\documentclass[letterpaper,10pt,conference]{ieeeconf}
\usepackage{graphicx}
\usepackage{color}

\IEEEoverridecommandlockouts                              

\usepackage{cite}
\usepackage{amsfonts,amsmath,amssymb}
\usepackage{enumerate}
\usepackage{comment}

\makeatletter
\let\NAT@parse\undefined
\makeatother
\usepackage{hyperref}%
\hypersetup{colorlinks=true, linkcolor=blue, breaklinks=true, urlcolor=blue, citecolor=blue}

\title{Opinion Dynamics with Set-Based Confidence:\\ Convergence Criteria and Periodic Solutions}

\author{Iryna Zabarianska and Anton V. Proskurnikov
\thanks{Iryna Zabarianska is with Department of Mathematics and Computer Science, St. Petersburg University.
Anton V. Proskurnikov is with Department of Electronics and Telecommunications at Politecnico di Torino, Turin.}
\thanks{Emails: \texttt{akshiira@yandex.ru,anton.p.1982@ieee.org}}
}

\def\be{\begin{equation}}
\def\ee{\end{equation}}
\def\ben{\begin{equation*}}
\def\een{\end{equation*}}

\newtheorem{example}{Example}
\newtheorem{theorem}{Theorem}
\newtheorem{lemma}{Lemma}
\newtheorem{assum}{Assumption}
\newtheorem{remark}{Remark}
\newtheorem{definition}{Definition}

\newcommand{\dfb}{\stackrel{\Delta}{=}}
\def\V{\mathcal V}
\def\E{\mathcal E}

\def\endofthm{\hfill$\star$}

\usepackage{subcaption}
\begin{document}

\maketitle

\begin{abstract}
This paper introduces a new multidimensional extension of the Hegselmann-Krause (HK) opinion dynamics model, where opinion proximity is not determined by a norm or metric. Instead, each agent trusts opinions within the Minkowski sum $\xi+\mathcal{O}$, where $\xi$ is the agent's current opinion and $\mathcal{O}$ is the confidence set defining acceptable deviations. During each iteration, agents update their opinions by simultaneously averaging the trusted opinions. Unlike traditional HK systems, where $\mathcal{O}$ is a ball in some norm, our model allows the confidence set to be non-convex and even unbounded.

We demonstrate that the new model, referred to as SCOD (Set-based Confidence Opinion Dynamics), can exhibit properties absent in the conventional HK model. Some solutions may converge to non-equilibrium points in the state space, while others oscillate periodically. These ``pathologies'' disappear if the set $\mathcal{O}$ is symmetric and contains zero in its interior: similar to the usual HK model, SCOD then converges in a finite number of iterations to one of the equilibrium points. The latter property is also preserved if one agent is "stubborn" and resists changing their opinion, yet still influences the others; however, two stubborn agents can lead to oscillations.
\end{abstract}

\section{Introduction}

The Hegselmann-Krause (HK) model~\cite{2002_hk} can be viewed as a deterministic averaging consensus algorithm with an opinion-dependent interaction graph, illustrating the principle of \emph{homophily} in social interactions: agents trust like-minded individuals and readily assimilate their opinions, while approaching dissimilar opinions with discretion. For historical discussions and an overview of the HK model's development over the past 20 years, refer to surveys~\cite{Hegselmann2023survey, liu2023survey, Bernardo2024survey}.

The original model from~\cite{2002_hk} addresses scalar opinions, but many opinions are better represented as vectors, capturing individuals' positions on multiple topics, like belief systems~\cite{FriedkinPro2016, 2017_tac_parsegov} or experts' assessments of multifaceted problems, such as probability distributions~\cite{Degroot_74} or resource allocation between multiple entities~\cite{friedkin2019mathematical}. This led to the development of multidimensional HK models~\cite{2012_nedic}, where opinion formation involves averaging opinions within a multidimensional ball centered on the agent's opinion, ignoring those outside. The key consideration is the norm used to measure the proximity of opinions, which is usually $\ell_2$ (Euclidean), 
$\ell_1$ (Manhattan)~\cite{douven2022network} or $\ell_{\infty}$~\cite{de2022multi}. The HK system with the Euclidean norm allows for convenient Lyapunov functions~\cite{2012_nedic,etesami2013termination} and a mechanical kinetic energy analogue employed in many convergence analyses~\cite{bhattacharyya2013convergence, etesami2019simple, 2017_chazelle}.

At the same time, there is no substantial experimental support for using the Euclidean or any specific norm to assess opinion proximity within the cognitive mechanisms underlying social homophily and social selection. Furthermore, as the dimension of the opinion space grows, the ``nearest-neighbor'' rules in opinion assimilation are undermined by the phenomenon of distance concentration, studied in data science~\cite{Beyer1999,Aggarwal2001,Radovanovic2009}, where distances between all pairs of points in high-dimensional random data tend to become equal. Using the $\ell_p$ distance, higher values of $p$ exacerbate this phenomenon. For instance, even in 2, 3, and 4 dimensions, $\ell_1$ norm outperforms the Euclidean norm in evaluating data similarity, but is surpassed by $\ell_p$ distances
\footnote{The $\ell_p$ metrics for $p<1$ is defined as $|x-y|_p \doteq \sum_i|x_i-y_i|^p$. This metrics is not associated to any norm, and the unit ball is non-convex.} 
with $p<1$~\cite{Aggarwal2001}.

\subsubsection*{Objectives} The goal of this work is to explore how much the properties of bounded confidence opinion dynamics depend on the distance-based homophily mechanism. To this end, we move away from distance-based confidence and examine a generalized model, termed SCOD (Set-based Confidence Opinion Dynamics), where the confidence ball is replaced by a \emph{set} of admissible opinion discrepancies, $\mathcal{O}$. An agent with opinion $\xi$ trusts opinions within the Minkowski sum $\xi + \mathcal{O}$, ignoring those outside; the averaging opinion update mechanism remains the same as in the HK model.

\subsubsection*{Contributions} We explore the properties of the SCOD system by identifying its similarities and differences with the standard HK model and examining the role of the set $\mathcal{O}$: 

\noindent\emph{(i)} The SCOD model inherits the HK model's convergence properties when $\mathcal{O}$ is symmetric and contains zero in its interior: the group splits into clusters with equal opinions, and the dynamics terminate after a finite number of stages.

\noindent\emph{(ii)} Under the same conditions as in \emph{(i)}, opinions remain convergent even with one stubborn agent who never changes their opinion but influences others. However, two stubborn agents can give rise to periodic oscillations.

\noindent\emph{(iii)} If these conditions on $\mathcal{O}$ are violated, the SCOD model can exhibit behaviors untypical for the HK model, e.g.,  
some solutions oscillate or converge to non-equilibrium points.

\subsubsection*{Structure of the paper} The SCOD model is introduced in Section~\ref{sec.model}, showing that even a small-size SCOD system with a general set $\mathcal{O}$ can behave very differently from the conventional HK model. In Section~\ref{sec.symm}, we formulate our main result, establishing the convergence of the SCOD in the case of \emph{symmetric} $\mathcal{O}$ and  (also in presence of identical stubborn agents). The proof of this theorem is given in Section~\ref{sec.proofs}. Section~\ref{sec.concl} concludes the paper.

\section{The Model Definition and Examples}\label{sec.model}

The SCOD model introduced below naturally extends the multidimensional HK model introduced in~\cite{2012_nedic}. 


\subsubsection{\bf Opinions}
Denote the set of agents by $\V$ and their number by\footnote{Hereinafter, the cardinality of a set $N$ is denoted by $|N|$.} $n=|\V|$. At period $t=0,1,\ldots,$
agent $i\in\V$ holds an opinion vector $\xi^{i}(t)\in\mathbb{R}^d$, whose element $\xi^i_k$ stands for the agent's position on topic $k\in\{1,\ldots,d\}$. 
The system's state is naturally written as the $n\times d$ matrix~\cite{Degroot_74,FriedkinPro2016,Bernardo2024survey} 
\[
\Xi(t)\dfb(\xi^i_k(t))^{i\in\V}_{k=1,\ldots,d.}
\]

\subsubsection{\bf Confidence graph} 
Each agent forms their opinions based on the ``similar'' opinions of their peers, with ``similarity'' relations defined by the \emph{confidence set} $\mathcal{O}\subseteq \mathbb{R}^d$ and conveniently characterized by a \emph{confidence graph} $\mathcal{G}(\Xi)=(\V,\E(\Xi))$. In this graph, the nodes represent the 
agents, and an arc $i\to j$ exists (agent $i$ trusts agent $j$'s opinion) if and only if $\xi^j-\xi^i\in\mathcal{O}$. 
Node $i\in\V$ has the set of (out-)neighbors
\begin{equation}\label{eq.neighbors}
 \mathcal{N}_i(\Xi)\dfb\{j\in\V: \xi^j \in \xi^i + \mathcal{O} \}.
\end{equation}
We adopt the following assumption, entailing that $i\in\mathcal{N}_i(\Xi)\,\forall i\in\V$ (i.e., each node has a self-loop). 
\begin{assum}[\bf Self-confidence]\label{asm.0}
$\mathbf{0}\in\mathcal{O}$.\endofthm
\end{assum}

\subsubsection{\bf The SCOD (Opinion Update Rule)} The mechanism of opinion evolution is same as in the HK Model. The opinion of agent $i$ is formed by averaging the trusted opinions,
\begin{equation}\label{eq.HK}
\xi^i(t+1)=\frac{1}{|\mathcal{N}_{i}(\Xi(t))|} \sum_{j \in \mathcal{N}_{i}(\Xi(t))}\xi^j(t),\quad i\in\V.
\end{equation}

\subsubsection{\bf Extension: Stubborn Agents}

The SCOD model can be generalized to include \emph{stubborn agents} whose opinions always remain unchanged.
The SCOD with a set of stubborn individuals $\V_{s}\subset\V$ and set of \emph{regular} agents $\V\setminus\V_s$
is the system~\eqref{eq.HK}, where $\mathcal{N}_i$ for regular agents $i\in\V\setminus\V_s$ is defined by~\eqref{eq.neighbors}, whereas $\mathcal{N}_i(\Xi)\equiv\{i\}\quad\forall i\in\V_s$.



\subsection{The SCOD vs. Previously Known Models} 
\begin{figure}[htb]
     \centering
     \begin{subfigure}[b]{0.4\columnwidth}
         \centering
         \includegraphics[width=\textwidth]{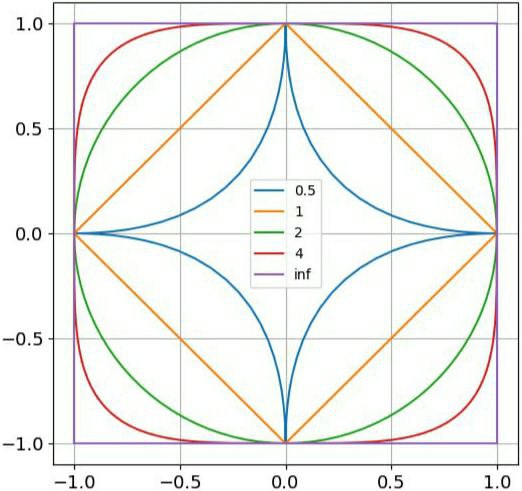}
         \caption{unit $\ell_{p}$-balls}
         \label{fig.balls}
     \end{subfigure}
     \hfill
     \begin{subfigure}[b]{0.38\columnwidth}
         \centering
         \includegraphics[width=\textwidth]{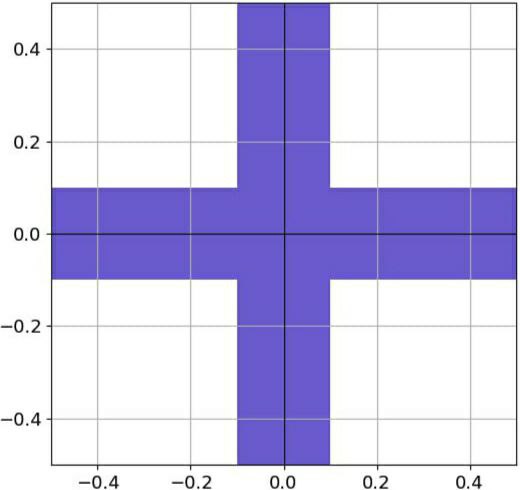}
         \caption{$\{\xi:\min\limits_{i=1,2}|\xi_i|\le 0.1\}$.}
         \label{fig.min_set}
     \end{subfigure}
    \caption{Examples of confidence sets.
    }
    \label{fig.conf_set}
\end{figure}

In the standard HK model the opinions are scalar ($d=1$), and $\mathcal{O}=(-R,R)$ is an interval\footnote{In some works~\cite{blondel2009krause}, closed intervals $[-R,R]$ have also been considered}. Later asymmetric intervals $\mathcal{O}=(-\ell,u)$ have been studied~\cite{2022_ECC_Bernardo}. 
Multidimensional HK models are special cases of the SCOD, where $\mathcal{O}$ is a ball centered at $0$0 with respect to some norm or metrics~\cite{2012_nedic,douven2022network,de2022multi}. Usually,
$\mathcal{O}$ is the $\ell_p$-ball (Fig.~\ref{fig.balls})
\[
\mathcal{O}=\mathcal{O}_{p,R}\dfb\left\{\xi\in\mathbb{R}^d:|\xi_1|^p+\ldots+|\xi_d|^p\leq R^p\right\}.
\]

Some models considered in the literature deal with unbounded confidence sets, e.g., the \emph{averaged-based} HK model from~\cite{de2022multi} is a special case of
~\eqref{eq.HK} with $\mathcal{O}=\{\xi\in\mathbb{R}^d:|\xi_1+\ldots+\xi_d|\leq R\}$ being a ``stripe'' between two hyperplanes.

Another interesting example is inspired by a more sophisticated dynamical model from~\cite{HuetDeffuant2008}. One may suppose that an agent with opinion vector $\xi$ can be influenced by another individual with opinion $\xi'$ if their positions $\xi_{k},\xi'_{k}$ on \emph{some} topic $k\in\{1,\ldots,d\}$ are close: $\mathcal{O}=\{\xi:|\xi_{l}|\leq\varepsilon_{l}\,\text{for some $k=1,\ldots,d$}\}$.
Fig.~\ref{fig.min_set} demonstrates this set for the special case of $d=2$ and $\varepsilon_1=\varepsilon_2=0.1$. 

\subsection{Gallery of Untypical Behaviors}

Before analyzing the general behavior of the SCOD system, we consider small-scale examples showing that with a general confidence set $\mathcal{O}$, it can behave very differently from standard HK models, where $\mathcal{O} = \{\xi : \|\xi\| \leq R\}$ is a ball. Namely, in the HK model (a) all solutions converge to equilibrium points in finite time, and (b) the agents split into clusters: those within a cluster reach consensus, while those in different clusters do not trust each other~\cite{proskurnikov2018tutorial}.
 None of these properties are generally valid for the SCOD model.

\subsubsection{\bf Non-clustered Equilibria}

The SCOD model can have equilibria, which are absent in the HK model. 
\begin{definition}\label{def.clustered}
Opinion matrix (the system state) $\Xi$ is \emph{clustered} if for all $i,j\in\V$ either $\xi^i=\xi^j$ or $\xi^j-\xi^i\not\in\mathcal{O}$.
\end{definition}

\begin{figure}[h]
\center
\includegraphics[width=3cm]{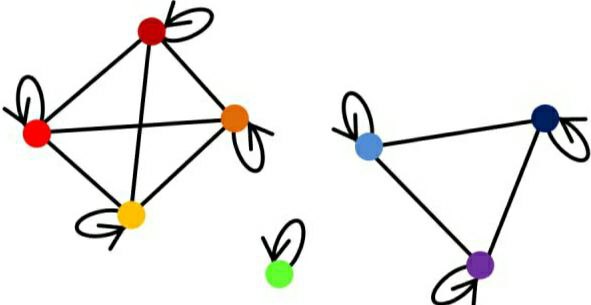}
  \caption{A union of disconnected cliques}
  \label{fig.cliques}
\end{figure}

A clustered matrix $\Xi$ is an equilibrium of the SCOD~\eqref{eq.HK}, and the graph $\mathcal{G}(\Xi)$ is a union of disjoint complete graphs, or \emph{cliques} (Fig.~\ref{fig.cliques}).
Unlike the HK model with norm-based confidence, SCOD systems admit \emph{non-clustered} equilibria.

\begin{figure}[htb]
     \centering     
     \begin{subfigure}[b]{0.24\columnwidth}
         \centering
         \includegraphics[width=\textwidth]{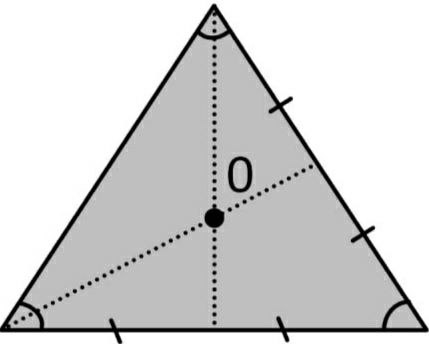}
         \caption{set $\mathcal{O}$}
        \label{fig.triangle_set}
    \end{subfigure}
    \hfill
     \begin{subfigure}[b]{0.27\columnwidth}
         \centering
         \includegraphics[width=\textwidth]{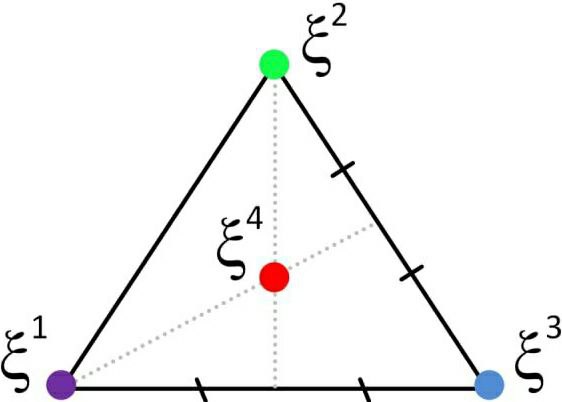}
         \caption{opinions}
        \label{fig.triangle_eqilib}
     \end{subfigure}
     \hfill
     \begin{subfigure}[b]{0.25\columnwidth}
         \centering
         \includegraphics[width=\textwidth]{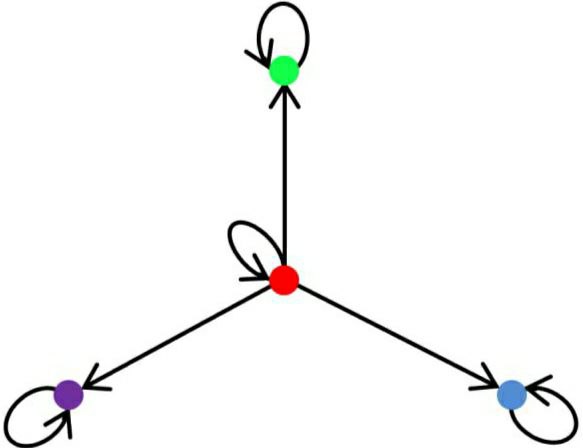}
         \caption{graph $\mathcal{G}(\Xi)$}
         \label{fig.triangle_graph}
     \end{subfigure}
    \caption{Non-clustered equilibrium of the SCOD}\label{fig.triangle}
\end{figure}

\begin{example}\label{ex.equi-patho1}
Choosing $\mathcal{O}$ as an equilateral triangle centered at the origin (Fig.~\ref{fig.triangle_set}) and choosing the opinions of $n=4$ agents as shown
in Fig.~\ref{fig.triangle_eqilib}, one gets an equilibrium of the SCOD that is not clustered as the strongly connected components of $\mathcal{G}(\Xi)$ are not disconnected (Fig.~\ref{fig.triangle_graph}).
\endofthm
\end{example}

\subsubsection{\bf Periodic Solutions} 
\begin{figure}[t]
\center
  \includegraphics[width=8cm]{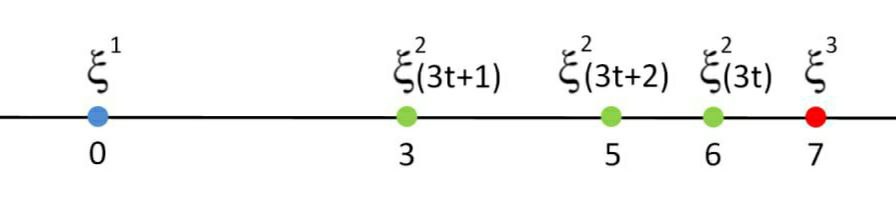}
  \caption{Example~\ref{ex.period_1d_n3}. $\xi^2$ oscillates with period 3.}
  \label{fig.3_period}
\end{figure}

We next show that small-size SCOD systems can exhibit periodic solutions.
\begin{example}\label{ex.period_1d_n3}
Consider $n=3$ agents and the confidence set
\begin{equation}\label{eq.setO-ex1}
\mathcal{O}=(-7,7)\setminus\mathcal{M},\;\mathcal{M}=\left\{\pm 1,\pm 3,\pm 5,-4,-2,6\right\}
\end{equation}
Then, the system~\eqref{eq.HK} has a periodic solution with $\xi^1\equiv 0$, $\xi^3\equiv 7$ (their sets of neighbors $\mathcal{N}_1\equiv\{1\}$, $\mathcal{N}_3\equiv\{3\}$ are constant) and
$\xi^2(t),\mathcal{N}_2(t)$ switching with period $3$ (Fig.~\ref{fig.3_period}):
\begin{equation}\label{eq.oscill1}
6
\xrightarrow[\mathcal{N}_2=\{1,2\}]{}3\xrightarrow[\mathcal{N}_2=\{2,3\}]{}5\xrightarrow[\mathcal{N}_2=\{2,3\}]{}6.
\end{equation}
\end{example}

\begin{remark} Notably, periodic solutions do not exist in the case where $\mathcal{O}$ is an interval, containing $0$~\cite{2015_acc_coulson,2022_ECC_Bernardo,Bernardo2024survey}; in this case
the dynamics terminate in time polynomially depending on $n$. Recent works~\cite{2020_vasca,bernardo2021heterogeneous}, focused on achieving of practical consensus under homophily and heterophily effects, also prove convergence in presence of a ``deadzone'' around $0$, in which case $\mathcal{O}=(-\ell,-\varepsilon)\cup\{0\}\cup(\varepsilon,u)$.\endofthm
\end{remark}

Our next example demonstrates that, when dealing with multidimensional opinions, periodic solutions are possible even with a confidence set being \emph{star-shaped} at $\mathbf{0}$. 
\begin{definition}
Set $\mathcal{O}$ is
\emph{star-shaped} at point $\xi^*$ if $[\xi^*,x]\dfb\{a\xi^*+(1-a)x:a\in[0,1]\}\subseteq\mathcal{O}$ for any $x\in\mathcal{O}$.  
For instance, a convex set is star-shaped at any of its points.
\endofthm
\end{definition}

If $\mathcal{O}$ is star-shaped at $\mathbf{0}$, then the following natural property holds. If an agent with opinion $\xi$ trusts another opinion $\xi'$, they trust all ``intermediate'' opinions from the interval $[\xi,\xi']$. 

\begin{figure}[htb]
     \centering
     \begin{subfigure}[b]{0.43\columnwidth}
         \centering
         \includegraphics[width=\textwidth]{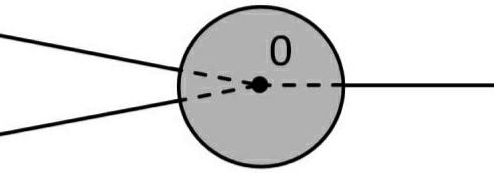}
         \caption{set $\mathcal{O}$}
         \label{fig.star_period_set}
     \end{subfigure}\hfill
     \begin{subfigure}[b]{0.5\columnwidth}
         \centering
         \includegraphics[width=\textwidth]{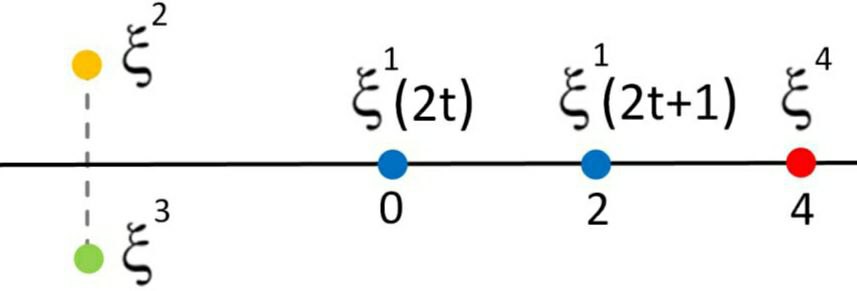}
         \caption{$n = 4$ opinions}
         \label{fig.star_period}
     \end{subfigure}
    \caption{Example~\ref{ex.period_2d}. (a) the confidence set; (b) opinions.}\label{fig.star_set_and_period}
\end{figure}

\begin{example}\label{ex.period_2d}
Consider a confidence set $\mathcal{O}\subset\mathbb{R}^2$ constituted by rays $\{\xi: \xi_1>0,\xi_2 = 0\}$, $\{\xi:\xi_2 = \xi_1/5<0\}$, $\{\xi:\xi_2 = -\xi_1/5>0\}$ and the unit circle (Fig.~\ref{fig.star_period_set}). Then,~\eqref{eq.HK} has a periodic solution (see Fig.~\ref{fig.star_period}) with $\xi^2\equiv (-3, 1), \xi^3\equiv (-3, -1), \xi^4\equiv (4, 0)$ and
$\xi^1(t),\mathcal{N}_1(t)$ switching as follows: 
\begin{equation}\label{eq.oscill2}
\xi^1=(0,0)\xrightarrow[\mathcal{N}_1=\{1,4\}]{} (2,0)\xrightarrow[\mathcal{N}_1=\{1,2,3,4\}]{}(0,0).
\end{equation}
\end{example}
\begin{remark}
In the latter example, unlike in Example~\ref{ex.period_1d_n3}, $\mathcal{O}$ is closed, but the periodic solution remains unchanged replacing $\mathcal{O}$ by its small open neighborhood.   \endofthm
\end{remark}

\begin{figure}[t]
\center
\includegraphics[width=4cm]{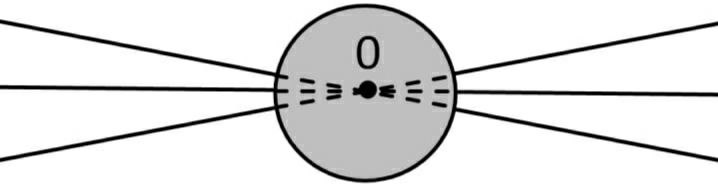}
  \caption{Example~\ref{ex.stubb}: set $\mathcal{O}$.}  \label{fig.stub_period}
\end{figure}

Revisiting Examples~\ref{ex.equi-patho1}-\ref{ex.period_2d}, an important feature is noted: the confidence set is asymmetric with respect to $\mathbf{0}$. This is not coincidental: as discussed in the next section, the symmetry ($\mathcal{O} = -\mathcal{O}$) \emph{excludes} the possibility of diverging solutions and non-clustered equilibria in the SCOD model without stubborn agents. However, the periodic solutions reemerge if the SCOD system with $\mathcal{O}=-\mathcal{O}$ includes stubborn agents ($\V_s\ne\emptyset$), as demonstrated by our next example.
\begin{example}\label{ex.stubb}
Consider a confidence set $\mathcal{O}$ which is a union of lines $\{\xi_2 = 0\}$, $\{\xi_2 = \xi_1/5\}$ and $\{\xi_2 = -\xi_1/5\}$ with the
ball of unit radius (see Fig.~\ref{fig.stub_period}). The SCOD with $n=4$, the set of stubborn agents $\V_s=\{2,3,4\}$ and the initial opinions from Fig.~\ref{fig.star_period} exhibits the oscillations in opinion $\xi^1$ as in~\eqref{eq.oscill2}.
Similarly, consider $n=3$ agents whose initial opinions are chosen as in Examples~\ref{ex.period_1d_n3}, but $\mathcal{O}=(-7,7)\setminus\{\pm 1,\pm 3,\pm 5\}$. If agents $1$, $3$ are stubborn, then
$\xi^2$ oscillates as in~\eqref{eq.oscill1}.\endofthm
\end{example}

\begin{remark}\label{rem.oscillations}
Note that in Examples~\ref{ex.period_1d_n3}-\ref{ex.stubb}, oscillations arise due to presence of static opinions, enabled by the geometry of set $\mathcal{O}$ or stubbornness of some agents. This effect, where static opinions induce oscillations, is well-known in models with randomized asynchronous interactions~\cite{AcemogluComo2013, Ravazzi2015}. Our examples show that the same effect occurs in the  \emph{deterministic} SCOD model with asymmetry or stubborn individuals\footnote{Notice that the systems in Examples~\ref{ex.period_1d_n3},\ref{ex.period_2d} are very different from their counterparts in Example~\ref{ex.stubb}, although the trajectories $\Xi(t)$ for the specific initial condition are same. In the former two examples, none of agents is stubborn, although some agents remain ``isolated'' ($\mathcal{N}_i\equiv\{i\}$) in the sense that they do not trust to the others because of the specific geometry of set $\mathcal{O}$ and the opinion trajectory $\Xi(t)$. In the latter example, some agents are stubborn and keep constant opinions \emph{for all possible initial conditions}.}.
\end{remark}

\subsubsection{\bf Convergent Solutions Absent in HK models}

\begin{figure}[htb]
     \centering
     \begin{subfigure}[b]{0.25\columnwidth}
         \centering
         \includegraphics[width=\textwidth]{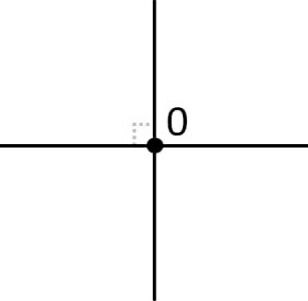}
         \caption{set $\mathcal{O}$}
         \label{fig.cross}
     \end{subfigure}
     \hfill
     \begin{subfigure}[b]{0.35\columnwidth}
         \centering
         \includegraphics[width=\textwidth]{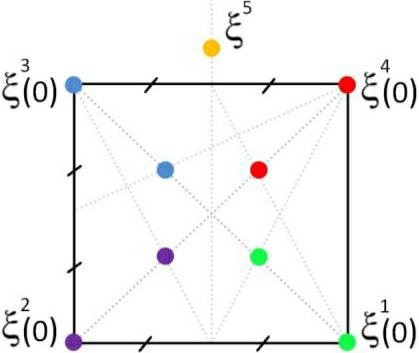}
         \caption{opinions}
         \label{fig.without_int}
     \end{subfigure}
    \caption{Example~\ref{ex.convergence-patho1}: (a) confidence set; (b) opinions.}\label{fig.conv-to-non-equi}
\end{figure}

Even if $\mathcal{O}$ is symmetric, solutions of the SCOD may converge in \emph{infinite} time and reach \emph{non-equilibrium} states\footnote{Similar behaviors are reported in continuous-time HK systems with generalized solutions~\cite{2012_narwa_ceragioli} yet are absent in the discrete-time HK model.}. This behavior is possible  as demonstrated by the following example.
\begin{example}\label{ex.convergence-patho1}
Let the two-dimensional confidence set be the union of two lines: $\xi_1 = 0$ and $\xi_2 = 0$ (Fig.~\ref{fig.cross}). The initial opinions of $n=5$ agents are shown in Fig.~\ref{fig.without_int}: four opinions are the vertices of the square $(\pm 1, \pm 1)$, while $\xi^5 = (0,a)$, where $a>1$. Evidently, $\xi^5$ is static, while $\xi^i$, $i=1,\ldots,4$ converge to $\mathbf{0}$. The resulting opinion profile is not an equilibrium. Removing the fifth agent, the solution converges over the \emph{infinite time} to the null equilibrium.\endofthm
\end{example}

\section{The SCOD with a Symmetric Confidence Set}\label{sec.symm}

Using the theory of averaging algorithms and inequalities~\cite{ProCalaCao}, it can be shown that for a \emph{symmetric} confidence set $\mathcal{O}=-\mathcal{O}$ the asymptotic behaviors of the SCOD model are similar to those of conventional HK models in the absence of stubborn agents. In the HK model based on the Euclidean norm, stubborn agents do not destroy convergence~\cite{2017_chazelle}, which, however, is not the case for the SCOD (see Example~\ref{ex.stubb}). Convergence can be guaranteed, however, in special situations, e.g., when only one agent is stubborn or all stubborn individuals share the same opinion. 

We first introduce the three key assumptions.
\begin{assum}[\bf Symmetric Confidence Set]\label{assum.Sym}
$\mathcal{O} = -\mathcal{O}$.\endofthm
\end{assum}
\begin{assum}[\bf Trust in Similar Opinions]\label{assum.Zero}
$\mathcal{O}$ contains $\mathbf{0}$ along with a small neighborhood\footnote{Since all norms on $\mathbb{R}^d$ are equivalent, the norm here is unimportant.}:
a radius $R>0$ exists such that $\mathcal{O}\supseteq\{\xi:\|\xi\|<R\}$.
\endofthm
\end{assum}

Assumption~\ref{assum.Sym} entails that the relations of trust are reciprocal: if $i$ trusts $j$, then $j$ trusts $i$ for each opinion matrix $\Xi$, in particular, graph $\mathcal{G}(\Xi)$ is undirected.
Assumption~\ref{assum.Zero} is a stronger form of Assumption~\ref{asm.0}, requiring the agent to trust all opinions that are sufficiently close (in the sense of usual distance) to their own.
For instance, the sets in Fig.~\ref{fig.conf_set} and Fig.~\ref{fig.stub_period} satisfy Assumptions~\ref{assum.Sym} and~\ref{assum.Zero}. The set in Fig.~\ref{fig.star_period_set} satisfies Assumption~\ref{assum.Zero} but violates Assumption~\ref{assum.Sym}, while the set in Fig.~\ref{fig.cross} satisfies Assumption~\ref{assum.Sym} but not Assumption~\ref{assum.Zero}.

\begin{assum}[\bf Homogeneous Stubborn Agents]\label{assum.Stub}
All stubborn agents (if they exist) share the same opinion\footnote{We assume that~\eqref{eq.stubb-same} holds automatically if $\V_s=\emptyset$.}:
\begin{equation}\label{eq.stubb-same}
\xi^i(0)\equiv\xi^*\quad\forall i\in\V_s.
\end{equation}
\end{assum}

\subsection*{Main Result: Convergence and Equilibria}

The following theorem examines the convergence of the SCOD trajectories $\Xi(t)$ and structures of their limits. 


\vspace{0.2cm}
\begin{theorem}\label{thm.converg_HK}
Assume that $\mathcal{O}$ obeys Assumptions~\ref{asm.0},~\ref{assum.Sym}, and $\Xi(0)$ obeys Assumption~\ref{assum.Stub}. The following statements are true:

 (A) $\Xi(0)$ is an equilibrium if and only if it is clustered.

 (B) All opinions have finite limits $\xi^i(\infty)=\lim_{t\to\infty}\xi^i(t)$, and $\xi^i(\infty)=\xi^j(\infty)$ whenever agents $i,j$ trust each other infinitely often
 $\xi^j(t_k)-\xi^i(t_k)\in\mathcal{O}$ for a sequence $t_k\to\infty$.
 
 \vskip0.1cm
\noindent If Assumption~\ref{assum.Zero} also holds, then:

 (C) The terminal state $\Xi(\infty)$ is a (clustered) equilibrium.

 (D) If $\V_s=\emptyset$ (no stubborn agents), the dynamics terminate in a finite number of steps. Otherwise, every opinion $\xi^i(t)$ either converges to the stubborn agents' common opinion $\xi^*$ from~\eqref{eq.stubb-same} 
 or stops changing after a finite number of steps.  \endofthm
\end{theorem}

\subsection{Numerical Example}

\begin{figure}[h]
     \centering
     \begin{subfigure}[b]{0.48\columnwidth}
         \centering
         \includegraphics[width=\textwidth]{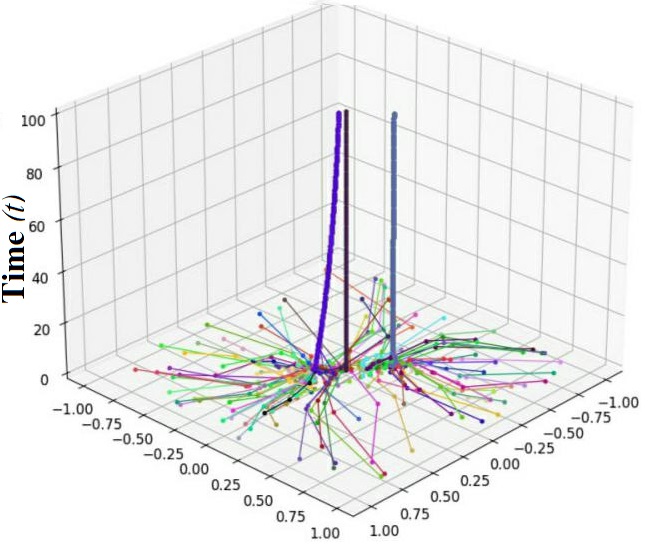}
     \end{subfigure}
     \hfill
     \begin{subfigure}[b]{0.48\columnwidth}
         \centering
         \includegraphics[width=\textwidth]{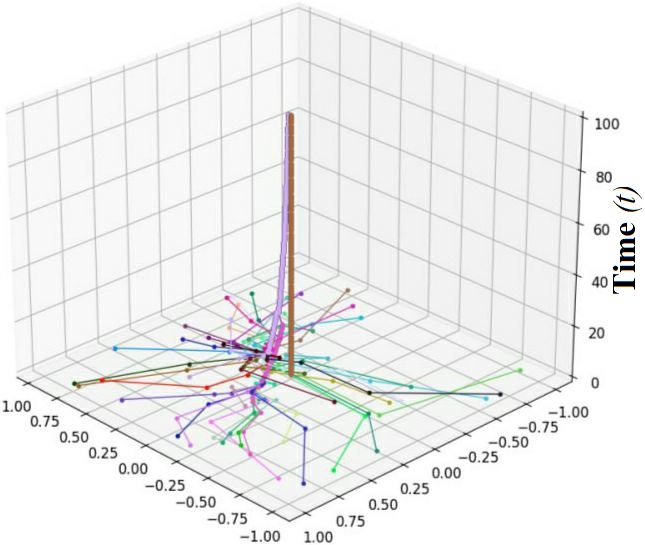}
     \end{subfigure}
    \caption{The SCOD with 1 (left) and 50 (right) stubborn agents} 
    \label{fig.scod-stubborn}
\end{figure}

The following numerical example illustrates the behavior of the SCOD with the set $\mathcal{O}$ from Fig.~\ref{fig.min_set} for $n=100$ agents and $\xi^*=\mathbf{0}$. The left plot in Fig.~\ref{fig.scod-stubborn} demonstrates the case where $|\V_s|=1$ and two clusters emerge. The right plot is for $|\V_s|=50$: the group reaches consensus at $\mathbf{0}$. The opinions of regular agents are sampled uniformly from $[-1,1]^2$. 

One may notice that the convergence to the stubborn opinion is quite slow; the estimate of the convergence rate in the SCOD models remains a non-trivial open problem.

\subsection{Discussion}

The assumptions of Theorem~\ref{thm.converg_HK}, while formally only sufficient, are essential and cannot be readily discarded.

\emph{Assumption~\ref{asm.0}}, besides making~\eqref{eq.HK} well-defined ($|\mathcal{N}_i| \ne \emptyset$), also excludes trivial periodicity due to infinite opinion swapping, e.g., if 
$\mathcal{O} = \mathbb{R}^d \setminus {0}$, the trivial SCOD dynamics $\xi^1(t+1) = \xi^2(t)$, $\xi^2(t+1) = \xi^1(t)$ violates (B). Furthermore, every pair of different opinion is a clustered state,
being, however, non-equilibrium, so (A) is also wrong.

Discarding \emph{Assumption~\ref{assum.Sym}}, even in the absence of stubborn agents, can result in oscillatory solutions (Examples~\ref{ex.period_1d_n3} and~\ref{ex.period_2d}). Even for converging solutions, $\Xi(\infty)$ need not be clustered: Example~\ref{ex.equi-patho1} shows that non-clustered equilibria may exist, even in the absence of stubborn agents. Hence, both (A) and (B) may be violated without the symmetry of $\mathcal{O}$.

\emph{Assumption~\ref{assum.Stub}} also cannot be fully discarded, as shown by Example~\ref{ex.stubb}: two stubborn agents with different opinions can lead to periodic solutions, even if $\mathcal{O}$ obeys Assumptions~\ref{asm.0}-\ref{assum.Zero}.

Notice that (A) \emph{does not} claim that the terminal state $\Xi(\infty)$ is an equilibrium. As Example~\ref{ex.convergence-patho1} shows, (C) is generally incorrect without \emph{Assumption~\ref{assum.Zero}}. The same example shows that a solution converging to an equilibrium need not reach it in finite time, so (D) can also be violated.

\section{Technical Proofs}\label{sec.proofs}


We will use the following lemma on the convergence of recurrent averaging inequalities~\cite[Theorem~5]{ProCalaCao}
\begin{gather}
x(t+1)\le W(t)x(t),\quad t=0,1\ldots,\label{eq.RAI}
\end{gather}
where $x(t)$ are $n$-dimensional column vectors, $W(t)$ are \emph{row-stochastic} $n\times n$ matrices and the inequality is elementwise.

\begin{lemma}\label{lm.consensus_alg}
Let matrices $W(t)$ be \emph{type-symmetric}, that is, for some constant $K\geq 1$ one has $K^{-1}w_{ji}(t)\leq w_{ij}(t) \leq Kw_{ji}(t)$ for all pairs $i\ne j$ and all $t=0,1,\ldots$
Assume also that the diagonal entries are uniformly positive: $w_{ii}(t)\geq \delta>0$ for all $i$ and $t\geq 0$. Then, any
solution $x(t)$ of~\eqref{eq.RAI} that is bounded from below enjoys the following properties:
\begin{enumerate}[(a)]
\item a finite limit $x(\infty) \dfb \lim_{t\rightarrow \infty}x(t)$ exists;
\item $x_i(\infty) = x_j(\infty)$ for all pairs of agents $i,j$ that interact \emph{persistently}, that is, $\sum_{t = 0}^{\infty}w_{ij}(t) = \infty$;
\item the residuals $\Delta(t)\dfb W(t)x(t)-x(t+1)$ are $\ell_1$-summable, that is, $\sum_{t=0}^{\infty}\Delta(t)<\infty$.\endofthm
\end{enumerate}
\end{lemma}
\begin{remark}
 Lemma~\ref{lm.consensus_alg} is well-known for averaging consensus algorithms \(x(t+1) = W(t)x(t)\), whose trajectories are always bounded from below and satisfy~\eqref{eq.RAI}. Under the assumptions of Lemma~\ref{lm.consensus_alg}, the consensus dynamics thus enjoys properties (a) and (b), with (c) being trivial. This statement, in a more general setting, appeared in~\cite[Theorem~1]{TouriCedric2014}, while its special case dates back to the seminal paper~\cite{2005_lorenz}.
\end{remark}

\subsection*{Case I: No Stubborn Agents}

Henceforth Assumptions~\ref{asm.0} and~\ref{assum.Sym} are supposed to be valid.

We first prove Theorem~\ref{thm.converg_HK} in the case where $\V_s=\emptyset$. The proof retraces one for the usual HK model~\cite{proskurnikov2018tutorial}. For a fixed solution $\Xi(t)$, the SCOD dynamics~\eqref{eq.HK} entails that
\begin{equation}\label{eq.averaging}
\xi^i(t+1)=\sum_{j\in\V}\bar w_{ij}(t)\xi^j(t),
\end{equation}
where matrices $\bar W(t)=(\bar w_{ij}(t))$ are determined by
\begin{equation}\label{eq.W_GHK}
\bar w_{ij}(t)\dfb 
\begin{cases}
  \frac{1}{|\mathcal{N}_{i}(t)|},  & \mbox{if } j \in \mathcal{N}_{i}(t)\\
  0, & \mbox{otherwise}
\end{cases}
\end{equation}
and satisfy the assumptions of Lemma~\ref{lm.consensus_alg} thanks to Assumptions~\ref{asm.0} and~\ref{assum.Sym}. Furthermore, $\bar w_{ij}(t)\in\{0\}\cup[1/n,\infty)$, and hence $i,j$ trust each other infinitely often if and only if
\begin{equation}\label{eq.aux1}
\sum_{t=0}^{\infty}\bar w_{ij}(t)=\infty.
\end{equation}

To prove (B), fix a coordinate \(k\in\{1, \ldots, d\}\). The vectors \(x(t) \dfb (\xi^1_k(t), \ldots, \xi^n_k(t))^{\top}\) obey the consensus dynamics \(x(t+1) = \bar W(t)x(t)\). Thus, the limits \(x(\infty) = \lim_{t \to \infty} x(t)\) exist, and \(x_i(\infty) = x_j(\infty)\) if~\eqref{eq.aux1} holds by virtue of Lemma~\ref{lm.consensus_alg}. 
Applying this for all \(k\), statement (B) follows.

To prove (A), notice that for an equilibrium $\Xi(t)\equiv\Xi$ the respective matrix $\bar W(t)\equiv \bar W$ is also constant. If two agents $i,j$ trust each other at the state $\Xi$, then $\xi^i=\xi^j$ in view of (B). This implies that every equilibrium is clustered (Definition~\ref{def.clustered}): agents $i,j$ cannot trust each other unless their opinions coincide.
Trivially, clustered states are equilibria.

Assume now that Assumption~\ref{assum.Zero} additionally holds. We will prove that the SCOD terminate in a finite number of steps, which implies both (C) and (D).
Notice first that $\mathcal{N}_i(t)\dfb\mathcal{N}_i(\Xi(t))=\{j:\xi^j(\infty)=\xi^i(\infty)\}$ for $t$ being large. 
Indeed, if $\lim_{t\rightarrow \infty}\xi^i(t) = \lim_{t\rightarrow \infty}\xi^j(t)$, then $\|\xi^i(t) - \xi^j(t)\|\leq R$ for $t$ being large, where $R$ is the radius from Assumption~\ref{assum.Zero}, whence $\xi^j(t) - \xi^i(t)\in\mathcal{O}$. On the other hand, we know that if $\xi^j(\infty)\ne\xi^i(\infty)$, then $j\not\in\mathcal{N}_i(t)$ starting from some step $t=t_{ij}$.
Hence, in a finite number of steps the graph $\mathcal{G}(\Xi(t))$ splits into several disconnected cliques (Fig.~\ref{fig.cliques}) and stops changing.
In view of~\eqref{eq.HK}, at the next step the agents in each clique reach consensus, arriving at an equilibrium. This finishes the proof of (C) and (D).

\subsection*{Case II: Stubborn Agent are Present}

Denote the set of ordinary agents by $\V'\dfb \V\setminus\V_s$. Without loss of generality, we assume that $\V'=\{1,\ldots,m\}$, where agents $\V_s=\{m+1,\ldots,n\}$. For each regular agent,
denote $x_i(t)\dfb\|\xi^i(t)-\xi^*\|$, where $\|\cdot\|$ is some norm on $\mathbb{R}^d$. 

\subsubsection*{\bf Step 1 - Recurrent Averaging Inequality} We first prove that vectors $x(t)=(x_1(t),\ldots,x_m(t))^{\top}$ satisfy inequality~\eqref{eq.RAI}, where the  stochastic matrices $W(t)$ are as follows
\begin{equation}\label{eq.W_GHK_stubb}
w_{ij}(t)\dfb 
\begin{cases}
\bar w_{ij}(t), &i,j\in\V',\,i\ne j,\\
\bar w_{ii}(t)+\sum\limits_{\ell\in\V_s'}\bar w_{i\ell}(t), & i=j\in\V',
\end{cases}
\end{equation}
where $\bar w_{ij}(t)$ are defined in~\eqref{eq.W_GHK}.
Indeed, fixing $i\in\V'$, using~\eqref{eq.averaging} and the norm's convexity, one arrives at
\[
x_i(t+1)\overset{\eqref{eq.averaging}}{=}\left\|\sum_{j\in\V}
\bar w_{ij}(t)(\xi^i(t)-\xi^*)
\right\|\leq \sum_{j\in\V}\bar w_{ij}(t)\|\xi^i(t)-\xi^*\|
\]
Note that the summand in the latter sum equals $\bar w_{ij}(t)x_j(t)$ when $j\in\V'$ and $0$ otherwise (because $\xi^j(t)\equiv\xi^j(0)=\xi^*$ for each stubborn agent $j\in\V_s$). Therefore,
\[
x_i(t+1)\leq \sum_{j\in\V'}\bar w_{ij}(t)x_j(t)\leq \sum_{j\in\V'} w_{ij}(t)x_j(t)\;\;\forall i\in\V'
\]
i.e.,~\eqref{eq.RAI} is satisfied. Furthermore, it is evident that
\begin{equation}\label{eq.delta-bound}
\begin{aligned}
\Delta_i(t)&\dfb \sum_{j\in\V}w_{ij}x_j(t)-x_i(t+1)\geq \\
&\geq (w_{ii}(t)-\bar w_{ii}(t))x_i(t)=x_i(t)\sum_{\ell\in\V_s}\bar w_{i\ell}(t).
\end{aligned}
\end{equation}

\subsubsection*{\bf Step 2 - Reduced-order SCOD} Matrices~\eqref{eq.W_GHK_stubb} satisfy the conditions of Lemma~\ref{lm.consensus_alg} thanks to Assumptions~\ref{asm.0} and~\ref{assum.Sym}. In view of Lemma~\ref{lm.consensus_alg}, the limit exists $x(\infty) \dfb\lim_{t\rightarrow \infty}x(t)$. Denote $I\dfb\{i\in\V':x_i(\infty)=0\}$ and $J\dfb\V'\setminus I$.

Recall that $x_i(\infty)=x_j(\infty)$ whenever agents $i,j\in\V'$ trust each other infinitely often, e.g.,~\eqref{eq.aux1} holds.
Hence, two agents $i\in I$ and $j\in J$ don't trust each other ($\bar w_{ij}(t)=0$) for $t$ being large.
Using statement (c) in Lemma~\ref{lm.consensus_alg} and the inequality~\eqref{eq.delta-bound}, one proves that
every agent $j\in J$ does not trust stubborn agents ($\bar w_{j\ell}(t)=0\,\forall\ell\in\V_s$) for $t$ being large.

For large $t$ the family of opinions $\tilde\Xi(t)=(\xi^j(t))_{j\in J}$ thus evolves \emph{independently} of the remaining group, following a \textbf{SCOD model of the reduced order $|J|$} without stubborn individuals.

\subsubsection*{\bf Step 3 - Reduction to Case~I} Statement (B) is now straightforward by noticing that the opinions of agents from $I\cup\V_s$ converge $\xi^*$, whereas the reduced-order SCOD converges in view of Case~I.
Furthermore, if agents $i,j$ trust each other infinitely often, then either $i,j\in J$ or $i,j\in I\cup\V_s$; in both cases 
the limit opinions coincide $\xi^i(\infty)=\xi^j(\infty)$.

To prove (A), consider an equilibrium solution $\Xi(t)\equiv\Xi$. Then, obviously, $\xi^i=\xi^*$ for $i\in I\cup\V_s$,
agents from $J$ do not trust to agents from $I$, and $\tilde\Xi=(\xi^j)_{j\in J}$ is an equilibrium of the reduced-order SCOD model, proved to be clustered. 
Hence, $\Xi$ is clustered; the inverse statement is obvious.

 If, additionally, Assumption~\ref{assum.Zero} holds, then the opinions from set $J$ stop changing $\tilde\Xi(t)=\tilde\Xi(\infty)$ for $t$ being large (statement (D) in Case~I), which proves (D) in the general situation. 
 We also know  that $\tilde\Xi(\infty)$ is a clustered state of the reduced-order SCOD,
 and agents $j\in J$ don't trust the stubborn individuals for $t$ being large, hence, $\xi^*-\xi^j(\infty)\not\in\mathcal{O}$.
 This proves that $\Xi(\infty)$ is also clustered, i.e., which finishes the proof of statement (C) and of our theorem. \hfill $\blacksquare$

\section{Conclusions and Open Problems}\label{sec.concl}

This paper extends the multidimensional Hegselmann-Krause model by replacing the distance-based opinion rejection mechanism with a general set-based mechanism. 
We analyze the resulting SCOD model, highlighting its similarities and differences with the usual (distance-based) HK model, and show that some properties of the HK model, such as finite-time convergence and equilibrium structure, extend to a symmetric confidence set containing 0 in its interior. 
However, this behavior can be disrupted by \emph{stubborn} individuals, whose presence may lead to periodic oscillations in the opinions of ``regular'' agents. Similar effects are well-known in stochastic gossip-based models (see Remark~\ref{rem.oscillations}) but, to the best of our knowledge, have not been captured by deterministic models.
Several examples in Section~\ref{sec.model} illustrate that for \emph{asymmetric} confidence set $\mathcal{O}$ the SCOD model behaves quite differently from the conventional HK models, exhibiting non-clustered equilibria, infinite-time convergence to non-equilibrium points and oscillatory trajectories. 

Finally, we mention several directions for future research.
\subsubsection*{\bf Stubborn Agents and Oscillations}
While Assumption~\ref{assum.Stub} cannot be discarded, it seems to be only sufficient for SCOD convergence. A natural question arises: when do stubborn agents give rise to oscillating trajectories?

\subsubsection*{\bf Convergence Rate}
A limitation of the averaging inequalities method~\cite{ProCalaCao} is the absence of explicit estimates on the convergence time or rate of the solutions. A natural question arises: how do the convergence time and rate in statement (D) of Theorem~\ref{thm.converg_HK} depend on $\mathcal{O}$ and $n$? 

\subsubsection*{\bf Heterogeneity and Attractors}
A natural extension of the SCOD model is the \emph{heterogeneous} SCOD, where each agent has its own confidence set $\mathcal{O}_i$. Stubborn agents can be naturally embedded into such a model by allowing $\mathcal{O}_i = \{\mathbf{0}\}$. Heterogeneous SCOD can have periodic solutions even if all $\mathcal{O}_i$ are open and symmetric\footnote{The case $n=3$ in Example~\ref{ex.stubb} can be modified by replacing stubborn agents with agents having small confidence intervals $\mathcal{O}_i=(-\varepsilon,\varepsilon)$.}.
On the other hand, heterogeneous HK models with balls of different radii are believed to converge~\cite{2012_SIAM_Mirtabatabaei,proskurnikov2018tutorial}, although a formal proof seems to be unavailable. This raises a natural question: under which assumptions does the heterogeneous SCOD model have periodic and other oscillatory solutions, and when do all its trajectories converge? Notice that this question is non-trivial even for the homogeneous SCOD studied in this paper.

\bibliographystyle{IEEEtran}
\bibliography{bib_from_survey,bibnew}

\end{document}